\documentclass[twocolumn,superscriptaddress,aps,preprintnumbers,amsmath,amssymb,sort&compress,nofootinbib
,a4paper
,prl
]{revtex4-2}
  
\usepackage{amsfonts,amsmath,amssymb,ascmac,bm,tensor}
\usepackage{fnpct} 
\usepackage{comment}
\usepackage{ifpdf}
\usepackage{slashed}
\usepackage{color}
\usepackage[mathscr]{eucal}
\usepackage{fourier}
\usepackage[utf8]{inputenc}
\usepackage{physics}
\usepackage{cancel}
\usepackage{subfig}

\usepackage{lipsum}
\ifpdf
  \usepackage{graphicx}     %   usepackage without driver option
  \usepackage[bookmarksopen,colorlinks=true,linkcolor=bblue,citecolor=bblue,urlcolor=ppink]{hyperref}
\else     % For (p)LaTeX + dvipdfmx
\fi

\usepackage{cancel}
\usepackage{ulem}
\usepackage{amsfonts}
\usepackage{amssymb}
\usepackage{graphicx}
\usepackage{amsmath,bm}
\usepackage{enumerate}
\usepackage{mathtools}
\usepackage{tikz} \usetikzlibrary{calc}
\usepackage{subfloat}
\usepackage{subfig}
\usepackage{xcolor}
\usepackage{float}
\usepackage{color}
\usepackage{here}

\usepackage{adjustbox}
\usepackage{graphicx,tikz}
\usetikzlibrary{mindmap}

\usepackage{mathrsfs}
\usepackage{float,epsfig}
\usepackage{dcolumn}% Align table columns on decimal point
\usepackage{pgfplots}
\usepackage{graphicx}% Include figure files
\usepackage{bm}% bold math
\usepackage{amsmath,amssymb,amsthm}

\usetikzlibrary{shadows}

\definecolor{red}{rgb}{1,0,0}
\definecolor{darkred}{rgb}{0.6,0,0}
\definecolor{darkgreen}{rgb}{0.992447,0.623778,0.034597}
\definecolor{ppink}{rgb}{1,0.4,0.4}
\definecolor{bblue}{rgb}{0.284602,0.317763,0.963947}
\definecolor{purple}{rgb}{0.5 ,0, 0.7}

\makeatletter
\newcommand\footnoteref[1]{\protected@xdef\@thefnmark{\ref{#1}}\@footnotemark}
\makeatother
 
\allowdisplaybreaks[1]

\definecolor{lime}{HTML}{A6CE39}
\newcommand{\orcidicon}{%
    \begin{tikzpicture}
    \draw[lime, fill=lime] (0,0)
        circle [radius=0.16]
        node[white] {{\fontfamily{qag}\selectfont \tiny ID}};
    \draw[white, fill=white] (-0.0625,0.095)
        circle [radius=0.007];
    \end{tikzpicture}   \hspace{-2mm}
}
\newcommand\orcidLahoucine{{\href{https://orcid.org/0000-0002-0143-5140}{\orcidicon}}}
\newcommand\orcidKarima{{\href{https://orcid.org/0000-0001-5419-8516}{\orcidicon}}}
\newcommand\orcidHasan{{\href{https://orcid.org/0000-0001-7408-0910}{\orcidicon}}}

\begin{document}

%%%%%%%%%%%%%%%%%%%%%%%%%%%
%%%%%%%%%%% Title %%%%%%%%%%%
%%%%%%%%%%%%%%%%%%%%%%%%%%%

\title{
% 1- A Supersymmetry fingerprint in accelerating black hole shadows
%  \\
%  2- ruled  by EHT observations\\
    A Supersymmetric Suspicion From Accelerating Black Hole Shadows 
    %Unveiling gravity's quantum fingerprint through
    }

\author{L. Chakhchi \orcidLahoucine}
\email{lahoucine.chakhchi@edu.uiz.ac.ma}
\author{H. El Moumni \orcidHasan}	
\email{h.elmoumni@uiz.ac.ma (Corresponding author)}
\affiliation{\small LPTHE, Physics Department, Faculty of Sciences,  Ibnou Zohr University, Agadir, Morocco.}

\author{K.  Masmar \orcidKarima.}
\email{karima.masmar@gmail.com}
\affiliation{\small LPTHE, Physics Department, Faculty of Sciences,  Ibnou Zohr University, Agadir, Morocco.}
\affiliation{\small Laboratory of  High Energy Physics and Condensed Matter
HASSAN II University, Faculty of Sciences Ain Chock, Casablanca, Morocco. }

\begin{abstract}
%Our study bridges the gap between astrophysics and theoretical physics, specifically in the realm of supersymmetry. This research 
%leverages groundbreaking observational data from the EHT, which captured the first-ever images of black hole shadows and offers compelling gravitational evidence for supersymmetry, which remains elusive in particle physics experiments. This could pave the way for new insights and directions in both fields.

In light of the Event Horizon Telescope (EHT) images of the supermassive black holes $\textrm{Sgr A}^\star$ and $\textrm{M87}^\star$, we explore a potential supersymmetry suspicion within the observational data. Specifically, we investigate the shadow of a supersymmetric accelerating black hole and compare our findings with observed quantities such as the angular diameter $\mathcal{D}$ and the fractional deviation $\bm{\delta}$. 
Our analysis reveals a significant alignment between the calculated quantities and the EHT collaboration measurements. This alignment suggests that the features of the black hole shadows observed by the EHT exhibit characteristics consistent with the supersymmetry framework.

Our results provide compelling evidence for supersymmetry from a gravitational perspective, which remains absent from the particle physics viewpoint till now.\\
{\bf Keywords:} 
Supersymmetry, EHT, Observational data,
Shadows, Accelerating black holes.

\end{abstract}

\date{\today}
\maketitle
%\preprint{CERN-TH-2020-182}
%\preprint{DESY 20-188}

%Over a century ago, Albert Einstein introduced his groundbreaking theory of gravitation, general relativity, which has since provided insights into various phenomena such as planetary orbits, starlight bending, cosmology, and the existence of objects like white dwarfs, neutron stars, and black holes. It was not until 2019 that humanity witnessed the first direct visual evidence of these enigmatic entities through the pioneering work of the Event Horizon Telescope (EHT) collaboration \cite{EventHorizonTelescope:2019dse}. The EHT successfully captured images of the supermassive black hole $\textrm{M87}^\star$ located at the core of the elliptical galaxy $M87$, and later in 2022, the black hole at the center of our own galaxy, known as $Sgr\ A^\star$ \cite{EventHorizonTelescope:2022urf}.

Over a century ago, Albert Einstein revolutionized our understanding of the universe by introducing his groundbreaking theory of gravitation, i.e. general relativity. This remarkable framework has since provided profound insights into a myriad of cosmic phenomena, shaping our comprehension of planetary orbits, the bending of starlight, the vast expanse of cosmology, and the existence of exotic celestial objects such as white dwarfs, neutron stars, and especially, black holes. 
The true marvel of Einstein's theory was not fully realized until 2019 when humanity bore witness to a historic moment—the first direct visual evidence of these enigmatic entities, thanks to the collaborative efforts of the Event Horizon Telescope (EHT) \cite{EventHorizonTelescope:2019dse}. Through the EHT's groundbreaking observations, we could gaze upon the awe-inspiring image of the supermassive black hole $\textrm{M87}^\star$ nestled at the heart of the elliptical galaxy $M87 $. 
Building upon this monumental achievement, the EHT continued to push the boundaries of astrophysical exploration. In 2022, the EHT once again captivated the world by capturing the elusive black hole residing at the center of our very own Milky Way galaxy, known as $\textrm{Sgr A}^\star$ \cite{EventHorizonTelescope:2022urf}. These extraordinary accomplishments not only validate Einstein's profound insights into the nature of spacetime but also open new frontiers in our quest to unravel the mysteries of the cosmos.

The remarkable images captured by the EHT have ushered in a new era in astrophysics and gravitational research. As visual evidence often speaks louder than equations alone, these images have propelled black holes into the spotlight, positioning them as invaluable laboratories for testing the principles of general relativity, as well as theories involving modified and quantum gravity. 
 Recent literature has extensively explored the implications of gravitational wave measurements for alternative theories of gravity with  extra dimensions of spacetime
 \cite{Banerjee:2019nnj,Chakraborty:2017qve,Chakravarti:2019aup}. Similarly, in-depth discussion exists regarding the shadow measurements from $\textrm{Sgr A}^\star$ and $\textrm{M}87^\star$,  by Tang {\it et al.} \cite{Tang:2022hsu}, where the black hole shadow observations obtained from the EHT were utilized to probe the existence of extra dimensions, leading to significant constraints on their size. These findings highlight the transformative impact of EHT's pictures on our understanding of the fundamental nature of the universe.
%\textcolor{red}{ref mann accel PRL \cite{Ashoorioon:2022zgu}}

 %These investigations provide valuable insights into the nature of gravity and the properties of such compact objects, further enriching our understanding of gravitation at the horizon scale.

It is widely understood that the geometry of spacetime is encapsulated in the metric tensor, in this sense, the original $C$ metric, which belongs to a class of exact solutions to Einstein field equations, was discovered by Hermann Weyl in 1917 \cite{Weyl}.
This solution can be understood as a pair of accelerating black holes apart from each \cite{Kinnersley:1970zw,Griffiths:2006tk}. Afterward in 1976, Plebaski-Demiaski's generalization expanded the $C$ black hole solutions to incorporate rotation and the cosmological constant \cite{Plebanski:1976gy}.
Investigations on the $C$ metric have a long history in literature, and the thermodynamics phase structure of the accelerating black holes has been formulated for small acceleration in AdS spacetime in recent years \cite{Appels:2016uha,Astorino:2016ybm,Appels:2017xoe,Anabalon:2018ydc}, with a special emphasis to the supersymmetric black hole case in \cite{Cassani:2021dwa}, where the authors identify a complex locus of supersymmetric and non-extremal solutions, defined through an analytic continuation of the parameters, upon which they obtain a simple expression for the on-shell action, while cognizant of that Supersymmetric black holes in AdS spacetime are inherently interesting  AdS/CFT correspondence despite the fact those solutions are rare finds \cite{Cacciatori:2009iz,Gnecchi:2013mta}.

In addition to gravity, our universe is governed by three other fundamental interactions, which are described by the Standard Model of particle physics. Supersymmetry (SUSY) offers a natural foundation for unifying gravity with such fundamental interactions and aids in the resolution of the high-energy physics hierarchy problem.
Furthermore,  gravitational backgrounds that preserve supersymmetry in supergravity theories are crucial to the development of string/M-theory, flux compactifications, and the AdS/CFT correspondence. 
Although one of the most promising frameworks for theories beyond the Standard Model, there is still a need to find its direct experimental support in the natural world.  In particular and from the particle physics perspective, since no significant excess in data is observed in CMS and ATLAS detectors in squarks searches in the proton-proton collisions at
$\sqrt{s}=13 TeV$, and where the exclusion limits at $95\%$ confidence level exclude top squark masses up to $1150 GeV$ \cite{ATLAS:2022ihe,CMS:2022tqr,CMS:2022uby,CMS:2022vpy}, the remaining possible hope for supersymmetry is the gravitational way.

%With all these motivations on our hands, and to explore the possible clues for supersymmetry theory from the Event Horizon Telescope (EHT) observations, we propose to study the shadow of the charged accelerating black hole with a cosmological constant and carrying ra rotation.

With a multitude of motivations driving our exploration and the potential for shedding light on the supersymmetry theory through observations by the Event Horizon Telescope (EHT), our proposal entails a comprehensive investigation into the shadow of a charged, accelerating black hole with a cosmological constant and possessing angular momentum.

The minimal $D=4$, $\mathcal{N}= 2$ gauged supergravity solution is described by  a bulk action given in \cite{Gauntlett:2001ur,Cassani:2021dwa} by
\begin{equation}\label{action}
S_{bulk}=\frac{1}{4\pi G_{(4)}}\int dx^4 \sqrt{-g}\left(R-\frac{6}{\ell^2}-F^2\right).
\end{equation}
In which  $F=dA$ represents the Maxwell field strength, while $\ell$ denotes the AdS (Anti-de Sitter) radius.  %and the cosmological constant $\Lambda$ is associated with $-3/\ell^2<0$. 
 The line element of the accelerating  black hole with a cosmological constant reads in Boyer–Lindquist coordinates as \cite{Cassani:2021dwa}
\begin{eqnarray}\nonumber
ds^{2}&=&\frac{1}{H^{2}}[-\frac{Q}{\Sigma}(\frac{1}{\kappa}dt-\chi d\phi)^{2}+\frac{\Sigma}{Q} dr^{2}+\frac{\Sigma}{P} d\theta^{2}\\&+&\frac{P}{\Sigma}(\frac{\chi}{\kappa} dt-(r^{2}+a^{2})\sin^{2}\theta d\phi)^{2}].
\end{eqnarray}
where the involved  functions are given by
\begin{eqnarray}\label{mmetric}
\nonumber
\chi &= &a\sin^{2}\theta,\ \Sigma(r,\theta) =r^{2}+a^{2}\cos^{2}\theta,\ H(r,\theta) =1-\mathcal{A} r \cos\theta,\\ 
P(\theta) &=&1-2\mathcal{A} m \cos\theta + (\mathcal{A}^{2}(a^{2}+e^{2}+g^{2})-\frac{a^{2}}{\ell^{2}})\cos^{2}\theta ,\\ \nonumber
Q(r) &=&(r^{2}-2mr+a^{2}+e^{2}+g^{2})(1-\mathcal{A}^{2}r^{2})+\frac{ r^{2}}{\ell^{2}}(a^{2}+r^{2}),\\ \nonumber
\end{eqnarray}
and the gauge field is obtained to be 
\begin{eqnarray}
A&=&A_tdt+A_\phi d\phi\\ \nonumber
&=&-e\frac{r}{\Sigma}\left(\frac{1}{\kappa}dt-a\sin^2\theta d\phi\right)+q\frac{\cos\theta}{\Sigma}\left(\frac{a}{\kappa}dt-(r^2+a^2)d\phi\right).
\end{eqnarray}
 Such a solution is parameterized by five quantities $m$, $e$, $g$, $a$ and $\mathcal{A}$, which stand for mass, electric charge, magnetic charge, angular momentum and acceleration parameter, respectively, additionally $\ell$ is the AdS radius and the constant $\kappa > 0$  is a trivial constant that can be absorbed in a rescaling of the time coordinate and help to normalized the Killing vector $\partial_t$ in order to get a first law of thermodynamics.

\begin{figure}[!ht]
 \centering
 \includegraphics[scale=.6]{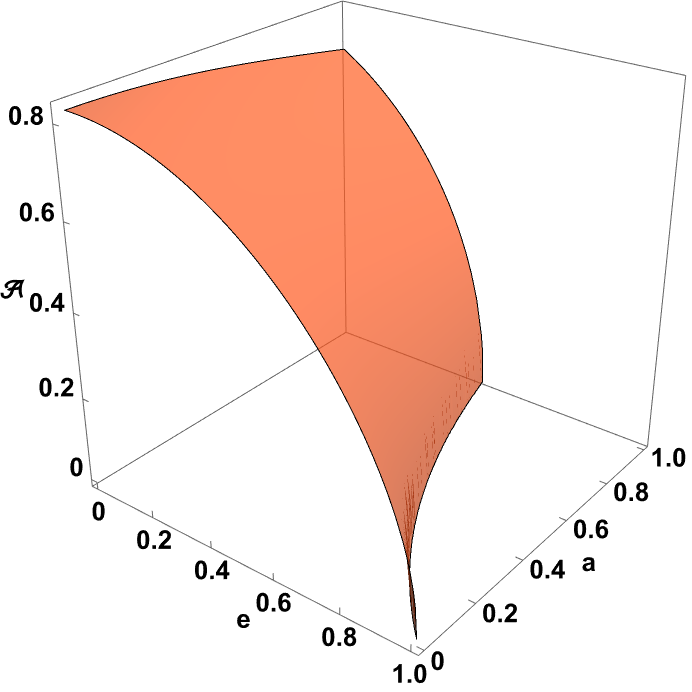}
  \includegraphics[scale=.6]{{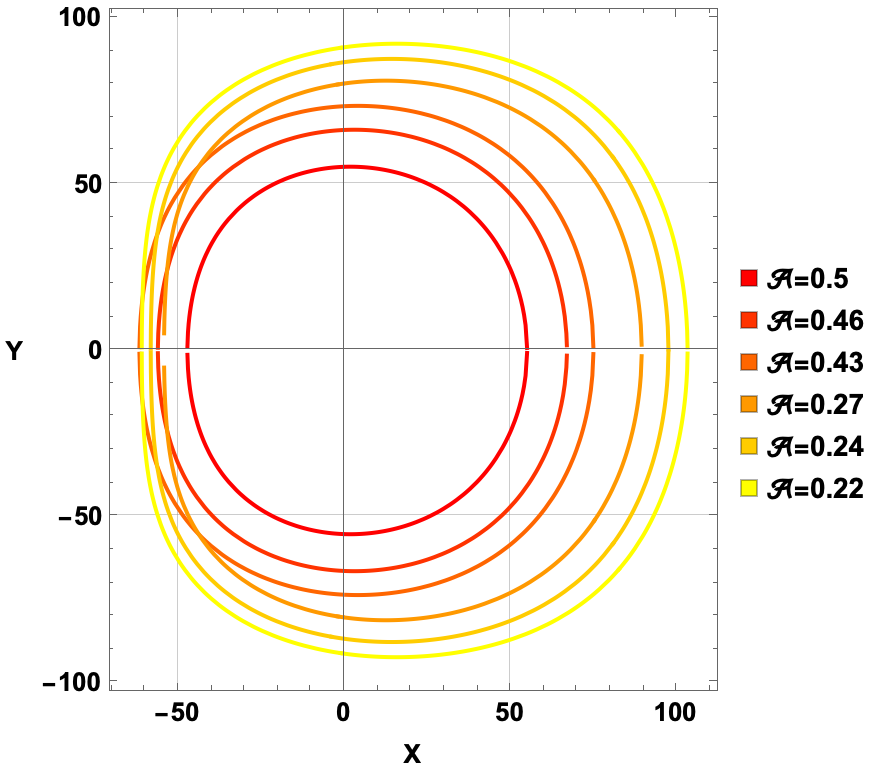}}
  \caption {\footnotesize\it  {\bf Top:} the supersymmetric moduli space associated with the constraints of Eq.\eqref{susycons}. 
  {\bf Bottom:} Apparent shape of the supersymmetric black hole, as seen by an observer at $(r_o=100,\theta_o= \frac{\pi}{2})$ for different values acceleration parameter $\mathcal{A}$. Here we have set the mass $m=1$. %\textcolor{red}{les axes $\alpha$ et $\beta$ noter l'acceleration autre que $\alpha$}
  }
\label{fig1}
\end{figure}

Adopting the strategy of \cite{Klemm:2013eca,Ferrero:2020twa}, it's possible to elaborate additional conditions on the parameters required for the solution to be supersymmetric based on Dirac spinor and Killing spinor equation. Indeed, for nonvanishing values of the accelerating parameter $\mathcal{A}\neq 0$ and for $\ell=1$, the constraints read as 
\cite{Cassani:2021dwa}  
\begin{eqnarray}\label{susycons}
g=\mathcal{A} m,\qquad
0=\mathcal{A}^2(e^2+g^2)(\Xi+a^2)-(g-a\mathcal{A} e)^2,
\end{eqnarray}
where, $\Xi\equiv 1+\mathcal{A}^2(a^2+e^2+g^2)- a^2$. 
Such constraints generate a $(e,a,\mathcal{A})$ moduli space depicted in the top panel of Fig.\ref{fig1}.
% In Fig.\ref{fig5} we depict such constraint in the $(a,\alpha,e)$ moduli space. It's also remarked that the supersymmetry constraint are independent of the cosmological constant $\Lambda$.

Exploring the shadow of such a black hole can unveil intricate details regarding the features of strong gravitational phenomena in the vicinity of the black hole horizon, potentially revealing a potential fingerprint of supersymmetry characteristics within the observational data obtained by the Event Horizon Telescope (EHT).

In this sense, studying the geodesics and orbits around the accelerating supersymmetric black hole involves the application of the Hamilton-Jacobi equations. By considering the tetrad components and the four-momentum, we can derive the coordinates describing the apparent displacement along both perpendicular and parallel axes to the projected axis of the black hole's symmetry. Utilizing a zero angular momentum observer (ZAMO) \cite{Johannsen:2013vgc} positioned at $(r_o,\theta_o)$, we parametrically determine the silhouette of the black hole shadow within the constraints of Eq.\eqref{susycons}, as illustrated in the bottom panel of Fig.\ref{fig1}.
 This figure reveals that the size of the  supersymmetric shadow cast decreases proportionally with the magnitude of the acceleration parameter $\mathcal{A}$. Furthermore, a precise consideration  of  the supersymmetric constraint in Eq.\eqref{susycons} unveils an additional influence on the shape of the black hole shadow by altering the rotation parameter $a$ and the black hole electric charge $e$ in conjunction with the acceleration $\mathcal{A}$. This interaction justifies the appearance of a D-shaped shadow for small values of $\mathcal{A}$.

It is inherent to anticipate that black hole parameters can be constrained through shadow analysis, as the shape and size of shadows directly correlate with these parameters. In seeking potential hints of supersymmetry, we turn our attention to key shadow observables, specifically the angular diameter and fractional deviation. By exploiting these quantities, we aim to unveil any indications of supersymmetric behavior. Subsequently, we juxtapose our derived results with observational data from the Event Horizon Telescope, focusing on measurements pertaining to $\textrm{M87}^\star$ and $\textrm{Sgr\ A}^\star$.

%Rigorously speaking, we consider in our investigation the angular diameter $\mathcal{D}$ of the black hole shadow, which is the key observable, and which is defined initially in \cite{Banerjee:2019nnj}. Besides, the EHT collaboration employed an additional quantity called the fractional deviation $\bm\delta$ to quantify the deviation between the infrared shadow radius and the Schwarzschild shadow one \cite{EventHorizonTelescope:2022apq} thus its utilization is a crucial task.  
In our investigation, we rigorously consider the angular diameter $\mathcal{D}$ of the black hole shadow as the key observable, initially defined in \cite{Banerjee:2019nnj}. Additionally, the Event Horizon Telescope (EHT) collaboration introduced the fractional deviation $\bm\delta$ to quantify the deviation between the infrared shadow radius and the Schwarzschild shadow radius \cite{EventHorizonTelescope:2022apq}. %\textcolor{red}{ The utilization of this fractional deviation is crucial for our next analysis}.

 According to reports from the Event Horizon Telescope (EHT) collaboration \cite{EventHorizonTelescope:2019pgp,EventHorizonTelescope:2019ggy}, the supermassive black hole $\textrm{M87}^\star$ at the center of the galaxy Messier 87 (M87) exhibits an angular diameter of $ \mathcal{D}=42 \pm 3 \mu as$, accompanied by a fractional deviation evaluated to $\bm{\delta}=-0.01_{-0.17}^{+0.17}$. Additionally, assuming the axis of rotation aligns with the jet axis, the inclination angle is estimated to be $17^{\circ}$. The collaboration also reports the source's mass as $(6.5\pm 0.7)\times 10^{9}M_{\odot}$ and its measured distance as $16.8\pm 0.8\textrm{Mpc}$.

Furthermore, the observed emission ring of $\textrm{Sgr\ A}^\star$, as reported by the Event Horizon Telescope collaboration \cite{EventHorizonTelescope:2022wkp}, has an angular diameter of $\mathcal{D}=51.8\pm 2.3 \mu as$. In contrast, the angular diameter of its shadow is estimated at $48.7\pm 7 \mu as$. Collaborative efforts, such as those from the Keck observatory team, provide estimates for the mass $m=(3.975\pm 0.058\pm 0.026)\times 10^{6} M_{\odot}$ and distance $d=(7959\pm59\pm32)\:\rm pc$ of $\textrm{Sgr\ A}^\star$ by leaving the red-shift parameter-free.
 Alternatively, the distance $d=(7935\pm50)\;\rm pc$ and mass $m=(3.951\pm 0.047)\times 10^{6} M_{\odot}$ are evaluated through assuming the red-shift parameter sets to unity \cite{Do:2019txf}.  
Similarly, collaborations involving the Very Large Telescope and the GRAVITY interferometer (VLTI) propose mass and distance estimates for $\textrm{Sgr\ A}^\star$, yielding values of $m=(4.261\pm0.012)\times 10^{6} M_{\odot}$ and $d=(8246.7\pm9.3)\rm pc$ \cite{GRAVITY:2021xju,GRAVITY:2020gka}. Adjustments for optical aberrations lead to slightly altered values of $m=(4.297\pm0.012\pm0.040)\times 10^{6} M_{\odot}$ and $d=8277\pm9\pm33\rm pc$. 
Otherwise, based on comparisons between the observed image of $\textrm{Sgr\ A}^\star$ and numerical simulation models, the inclination angle $i$ is inferred to exceed $50^\circ$. A specific value of $i\simeq 134^\circ$ (or equivalently $46^\circ$) is adopted to calculate the theoretical angular diameter of $\textrm{Sgr\ A}^\star$ \cite{refId0}. Additionally, the fractional deviation $\bm{\delta}$ obtained using the \textit{eht-img} algorithm \cite{EventHorizonTelescope:2022xqj} is ${\bm\delta}=-0.08_{-0.09}^{+0.09}$ according to VLTI measurements,  and ${\bm\delta}=-0.04_{-0.10}^{+0.09}$ within the Keck estimations.

\begin{figure*}
\centering
\hspace{-2cm}
\begin{tikzpicture}

\draw[olive, ultra thick, fill=olive!6] (3.8,17.3) rectangle (10.2,-1);
\draw[orange!20, ultra thick, fill=orange!10] (-1.8,17.3) rectangle (-8.2,-1);
\draw[orange!60, ultra thick, fill=orange!20] (-8.7,-1.4) rectangle (10.7,-7.6);
\draw[orange!60, ultra thick, fill=orange!20] (-8.7,-1.2) rectangle (-2.35,-7.8);
\draw[orange, ultra thick,fill=orange!10] (-1,8) -- (3,4) -- (-1,4) -- cycle;
\draw[olive, ultra thick, fill=olive!6] (3,8) -- (3,4) -- (-1,8) -- cycle;
\filldraw (1.,6) circle [white,radius=0.5, fill=red] node {\textcolor{white}{\Large$\bm{\mathcal{D}}$}};
 \node[text width=3cm] at (3,7)     {\Large$\textrm{M87}^\star$};
  \node[text width=3cm] at (1,5)     {\Large$\textrm{Sgr\ A}^\star$};
\draw[->,orange!60,double distance=4pt] (2,4)--(2,-1.32);
\draw[-,orange!20,double distance=4pt] (0,4)--(0,2);
\draw[->,orange!20,double distance=4pt] (0.,2)--(-1.7,2);
\draw[-,olive,double distance=4pt] (1,8)--(1,10);
\draw[->,olive,double distance=4pt] (.93 ,9.916)--(3.7,9.916);

%\node at (.3,16.7)  {\Large$\mathcal{D}$ \usebox0}; 
\node at (0,2.05) [rectangle,draw,fill=orange!10,node contents={\large Keck}];   
\node at (2,0.5) [rectangle,draw,fill=orange!20,node contents={\large VLTI}];       
    
\node[inner sep=0pt, text width=4cm ]  at (6,14.15)
    {\includegraphics[width=1.5\textwidth]{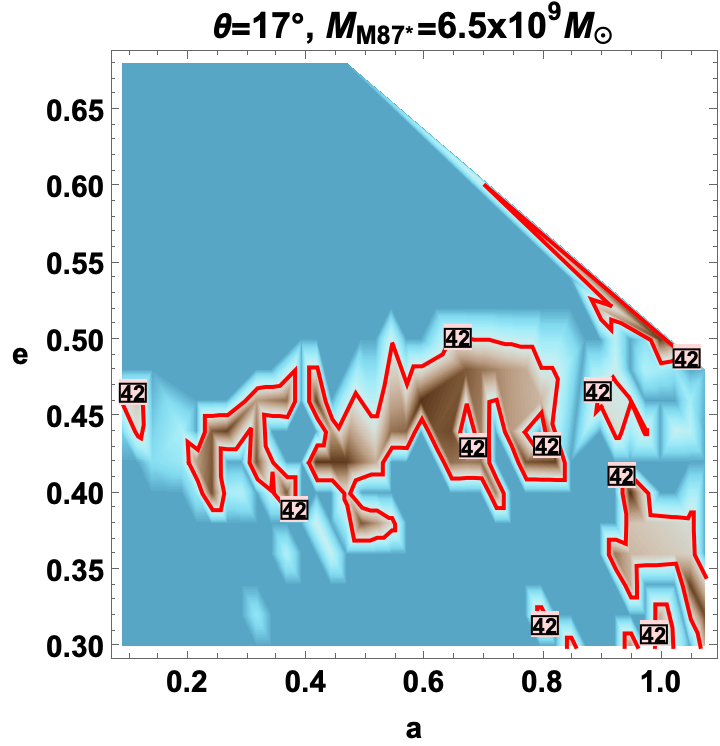}};
    \node[inner sep=0pt, text width=4cm ]  at (6,8)
    {\includegraphics[width=1.5\textwidth]{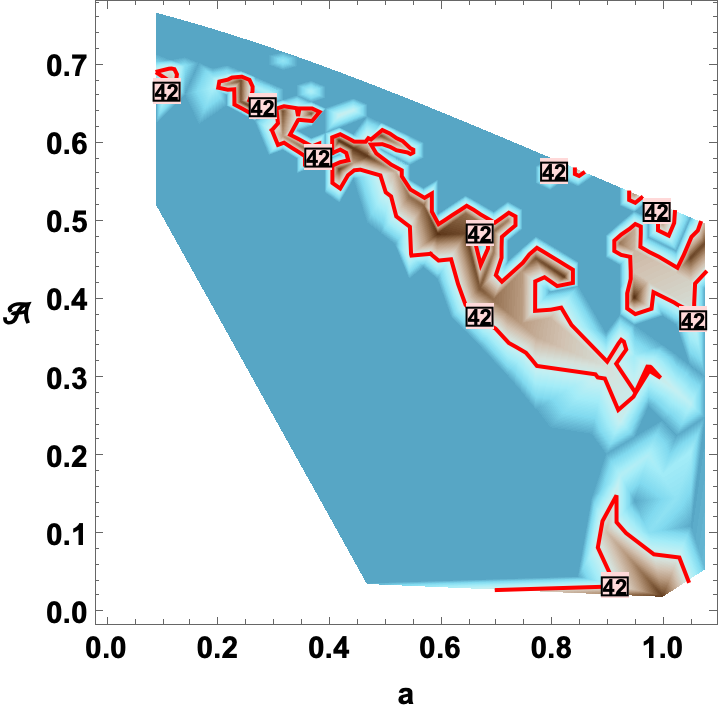}};
     \node[inner sep=0pt, text width=4cm ]  at (6,2)
    {\includegraphics[width=1.5\textwidth]{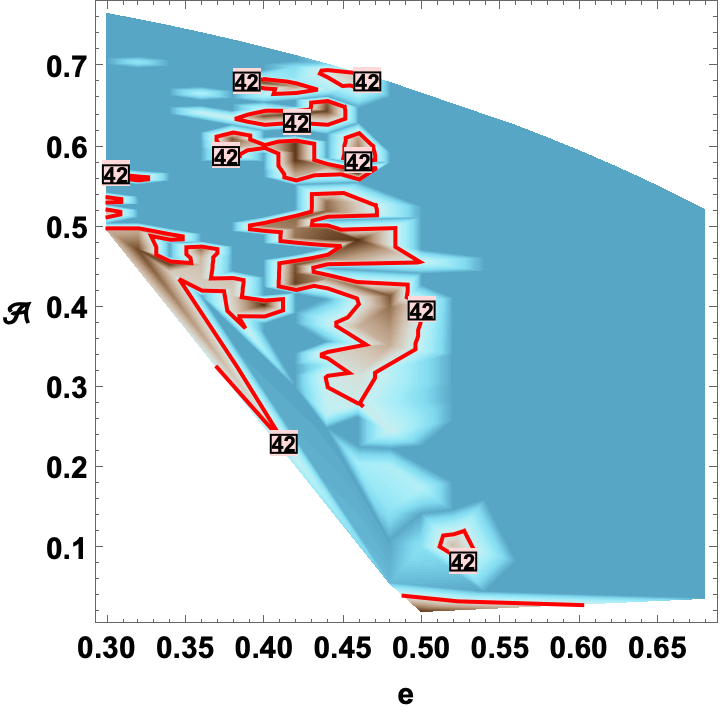}};

    \node[inner sep=0pt, text width=4cm]  at (6.5,-4.5)
    {\includegraphics[width=1.5\textwidth]{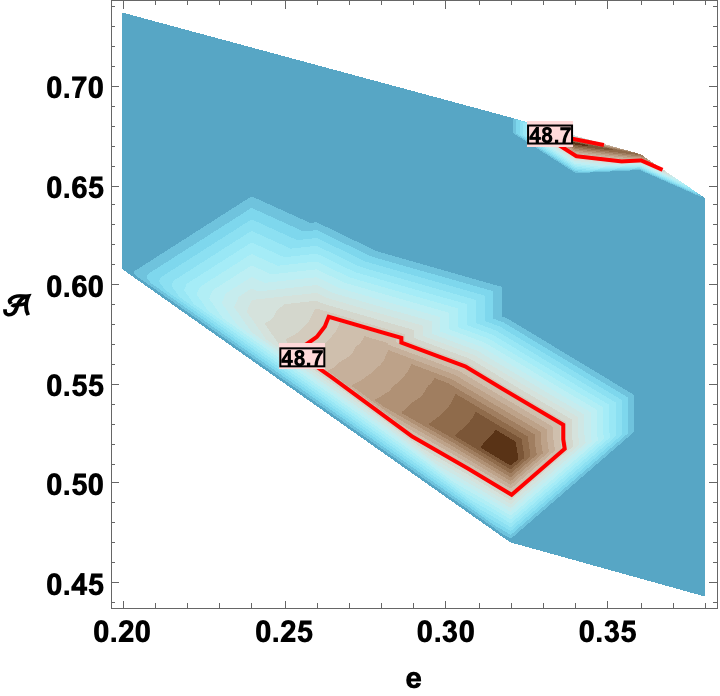}};
    \node[inner sep=0pt, text width=4cm ]  at (0,-4.5)
    {\includegraphics[width=1.5\textwidth]{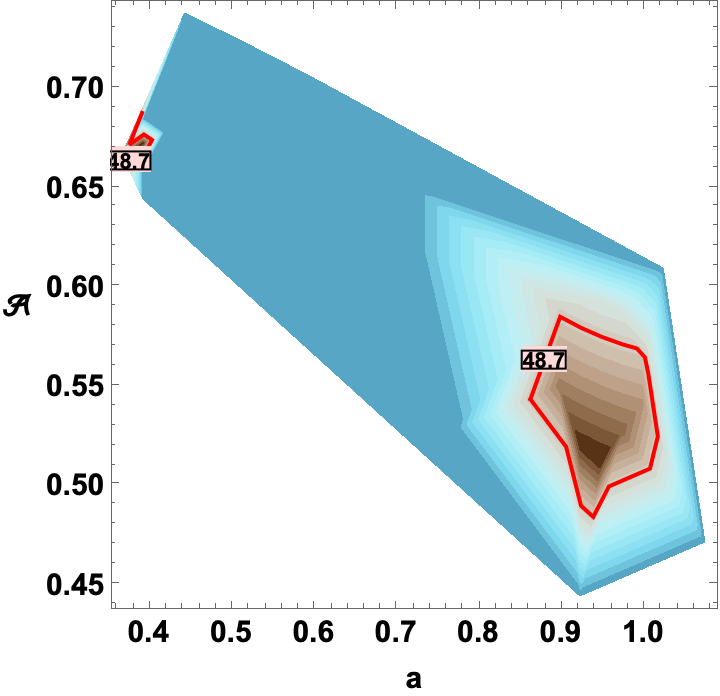}};
     \node[inner sep=0pt, text width=4cm ]  at (-6.5,-4.5)
    {\includegraphics[width=1.5\textwidth]{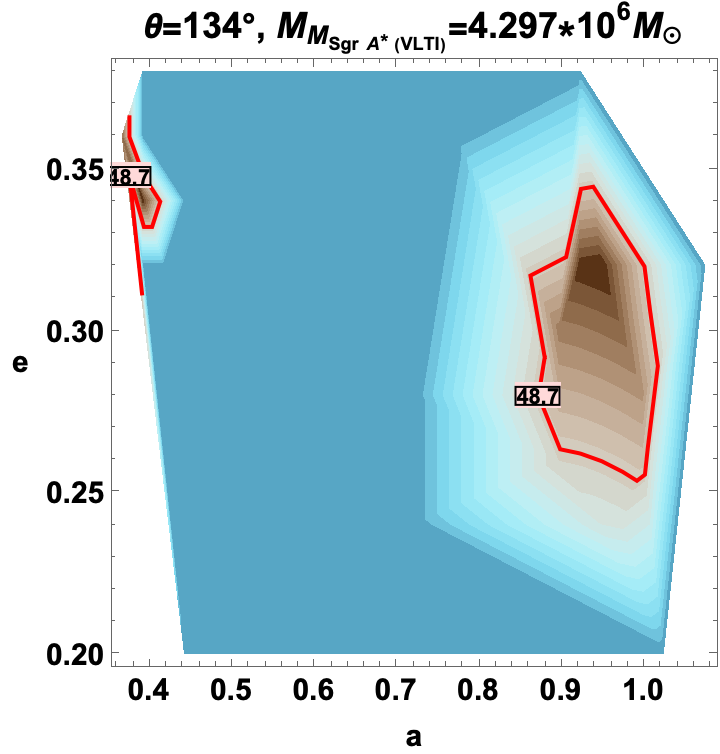}}; 
    
    \node[inner sep=0pt, text width=4cm ]  at (-6,14)
    {\includegraphics[width=1.5\textwidth]{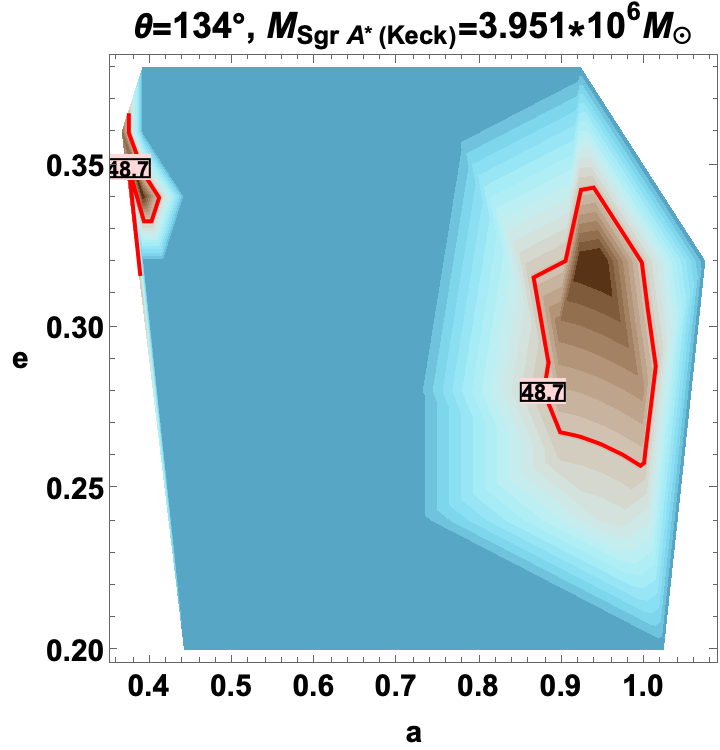}};
    \node[inner sep=0pt, text width=4cm ]  at (-6,8)
    {\includegraphics[width=1.5\textwidth]{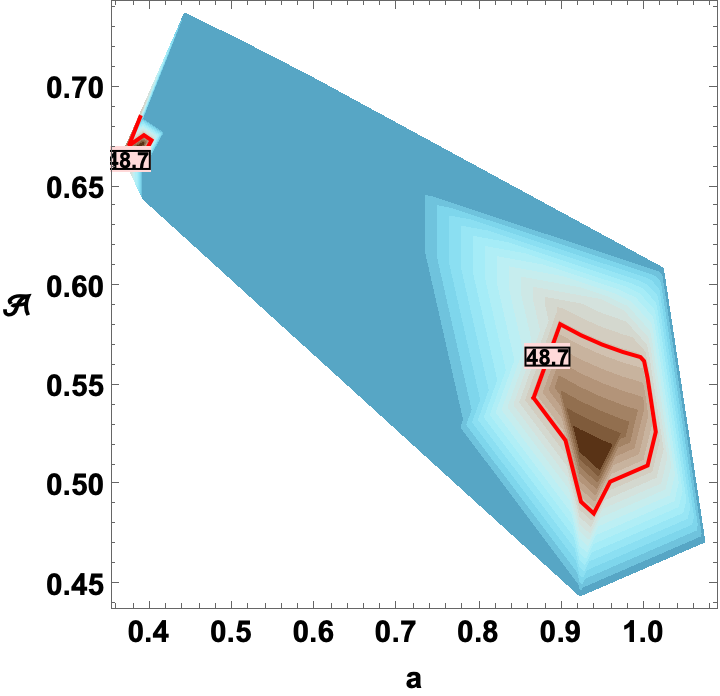}};
     \node[inner sep=0pt, text width=4cm ]  at (-6,2)
    {\includegraphics[width=1.5\textwidth]{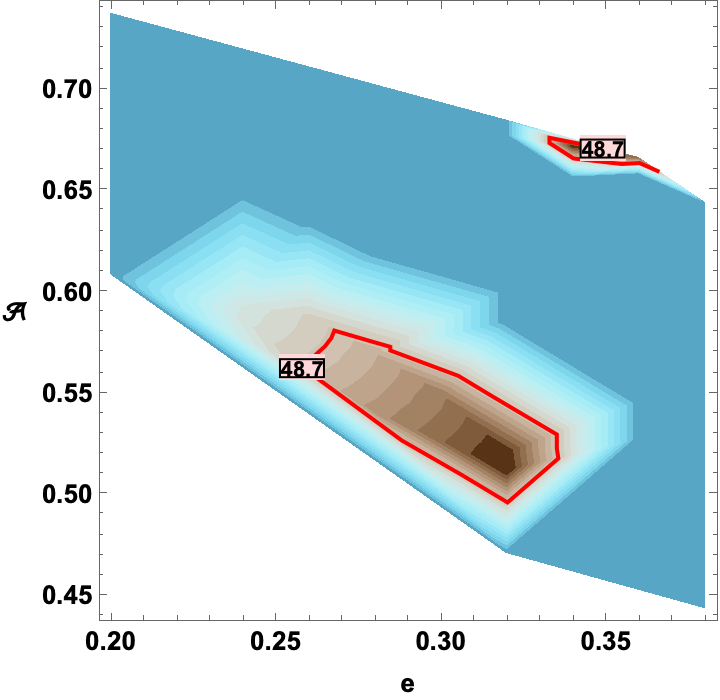}}; 

     \node[inner sep=0pt, text width=4cm ]  at (4,13.6) {\includegraphics[width=.4\textwidth]{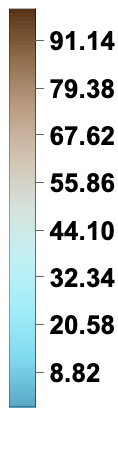}};
 \node[inner sep=0pt, text width=4cm ]  at (.5,13.3) {\includegraphics[width=.4\textwidth]{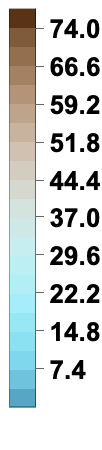}};
  \node[inner sep=0pt, text width=4cm ]  at (-1,-8.4) {\includegraphics[width=2\textwidth]{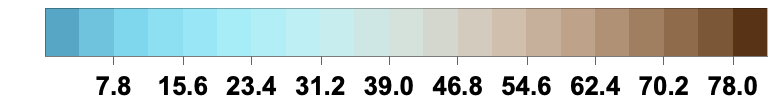}};

\end{tikzpicture}
\caption{\footnotesize  \it The contours illustrate the dependence of the angular diameter $\mathcal{D}$ for $\textrm{M87}^\star$ and $\textrm{Sgr~A}^\star$ (VLTI \& Keck observatories) on different planes of the supersymmetric accelerating black hole parameters, namely charge $e$, rotation $a$, and acceleration $\mathcal{A}$. The solid red line represents the EHT measurements.
}
\label{fig2}
\end{figure*}
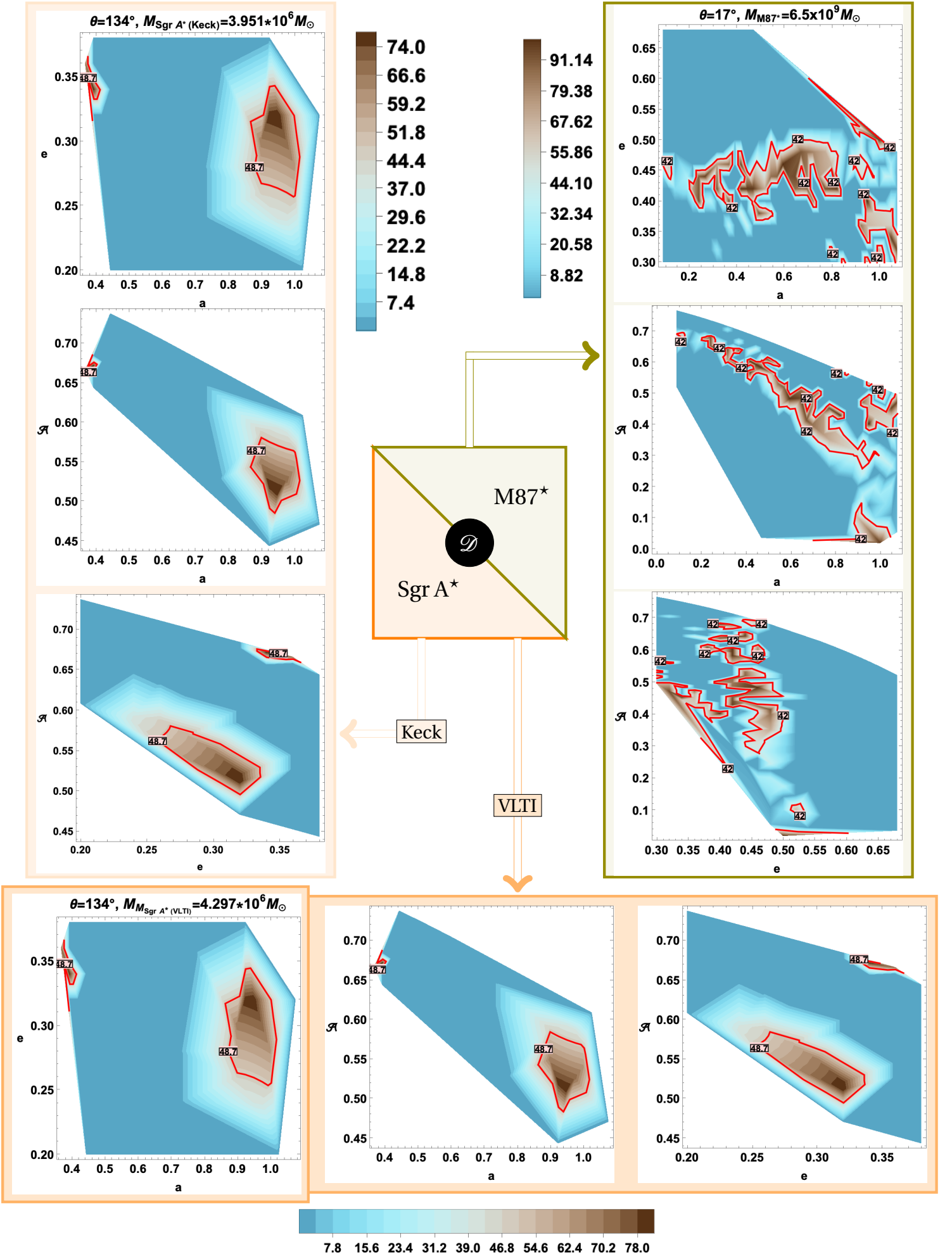

\begin{figure*}
\centering
\hspace{-2cm}
\begin{tikzpicture}

\draw[olive, ultra thick, fill=olive!6] (3.8,17.3) rectangle (10.2,-1);
\draw[orange!20, ultra thick, fill=orange!10] (-1.8,17.3) rectangle (-8.2,-1);
\draw[orange!60, ultra thick, fill=orange!20] (-8.7,-1.4) rectangle (10.7,-7.6);
\draw[orange!60, ultra thick, fill=orange!20] (-8.7,-1.2) rectangle (-2.35,-7.8);
\draw[orange, ultra thick,fill=orange!10] (-1,8) -- (3,4) -- (-1,4) -- cycle;
\draw[olive, ultra thick, fill=olive!6] (3,8) -- (3,4) -- (-1,8) -- cycle;
\filldraw (1.,6) circle [white,radius=0.5, fill=red] node {\textcolor{white}{\Large$\bm{\bm{\delta}}$}};
 \node[text width=3cm] at (3,7)     {\Large$\textrm{M87}^\star$};
  \node[text width=3cm] at (1,5)     {\Large$\textrm{Sgr\ A}^\star$};
\draw[->,orange!60,double distance=4pt] (2,4)--(2,-1.32);
\draw[-,orange!20,double distance=4pt] (0,4)--(0,2);
\draw[->,orange!20,double distance=4pt] (0.,2)--(-1.7,2);
\draw[-,olive,double distance=4pt] (1,8)--(1,10);
\draw[->,olive,double distance=4pt] (.93 ,9.916)--(3.7,9.916);

%\node at (.3,16.7)  {\Large$\mathcal{D}$ \usebox0}; 
\node at (0,2.05) [rectangle,draw,fill=orange!10,node contents={\large Keck}];   
\node at (2,0.5) [rectangle,draw,fill=orange!20,node contents={\large VLTI}];       
    
\node[inner sep=0pt, text width=4cm ]  at (6,14.15)
    {\includegraphics[width=1.5\textwidth]{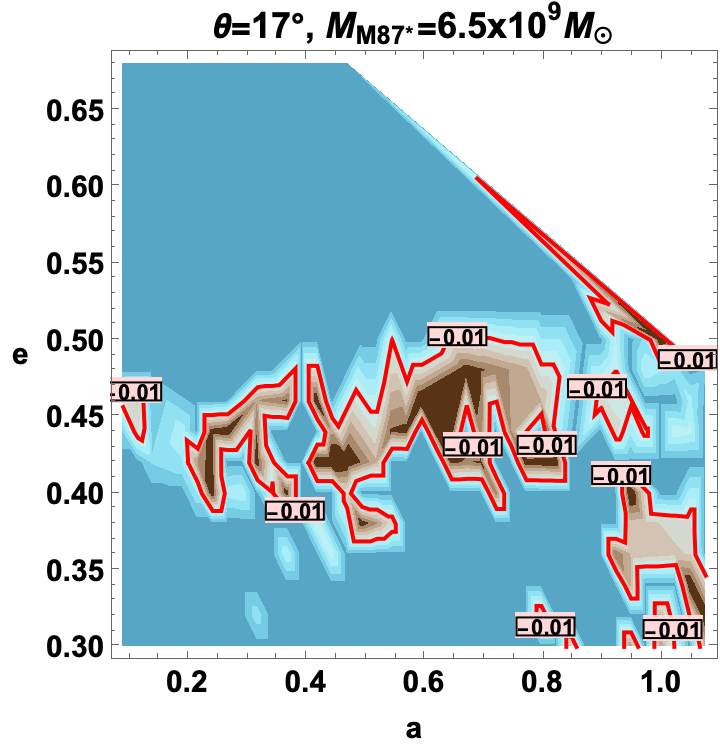}};
    \node[inner sep=0pt, text width=4cm ]  at (6,8)
    {\includegraphics[width=1.5\textwidth]{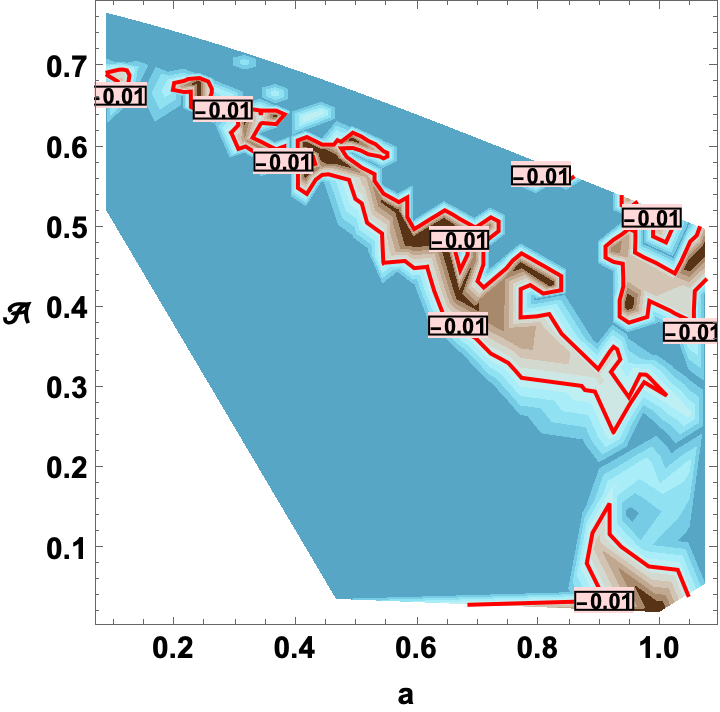}};
     \node[inner sep=0pt, text width=4cm ]  at (6,2)
    {\includegraphics[width=1.5\textwidth]{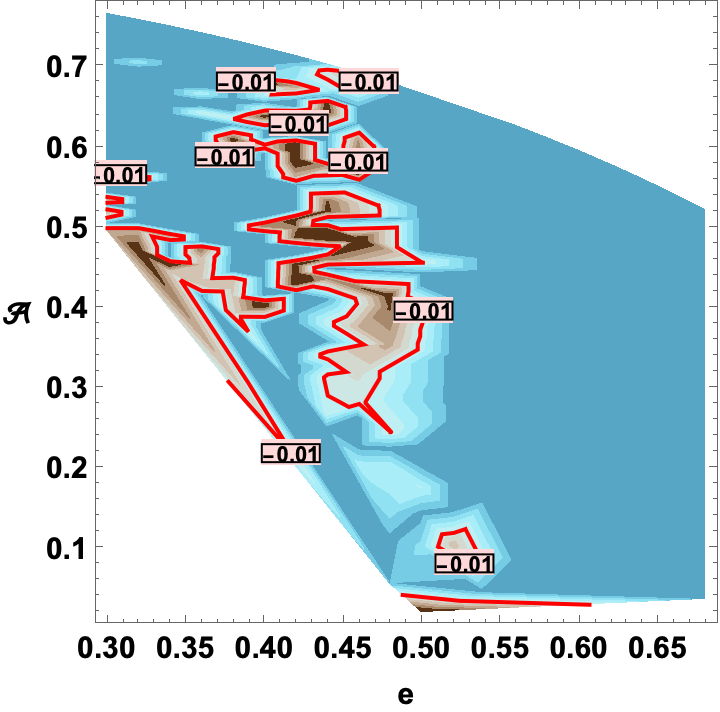}};

    \node[inner sep=0pt, text width=4cm]  at (6.5,-4.5)
    {\includegraphics[width=1.5\textwidth]{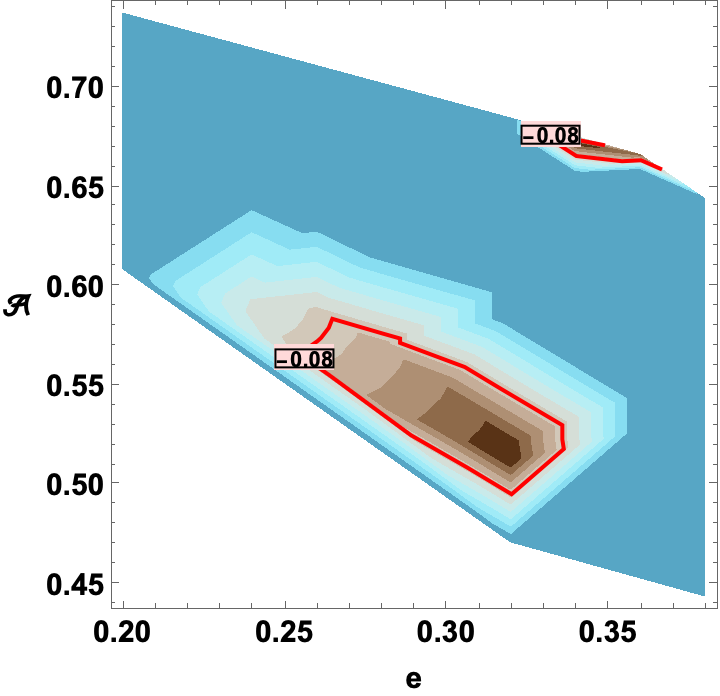}};
    \node[inner sep=0pt, text width=4cm ]  at (0,-4.5)
    {\includegraphics[width=1.5\textwidth]{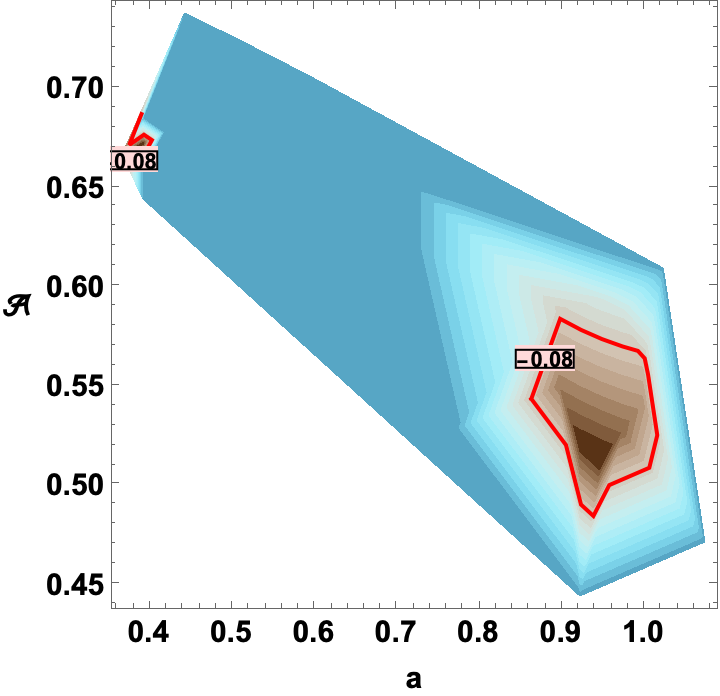}};
     \node[inner sep=0pt, text width=4cm ]  at (-6.5,-4.5)
    {\includegraphics[width=1.5\textwidth]{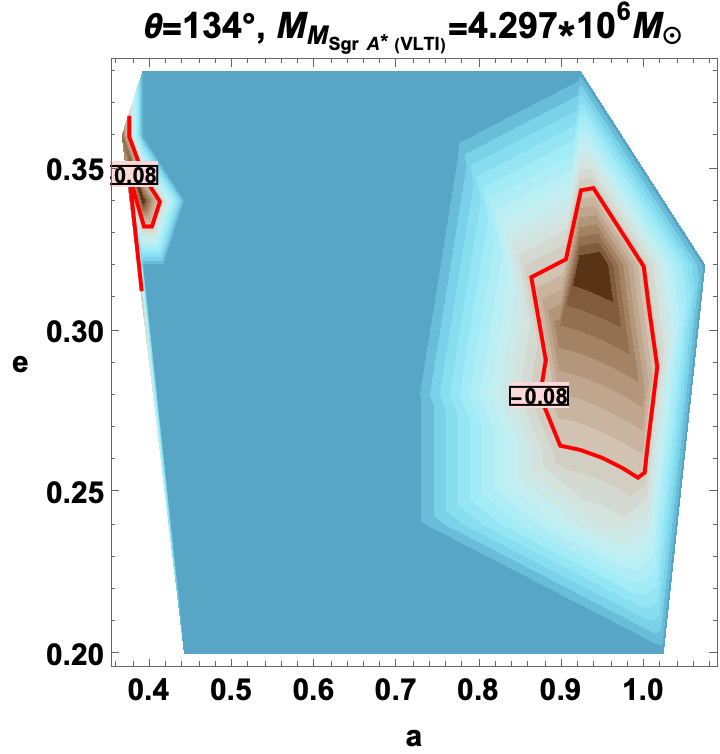}}; 
    
    \node[inner sep=0pt, text width=4cm ]  at (-6,14)
    {\includegraphics[width=1.5\textwidth]{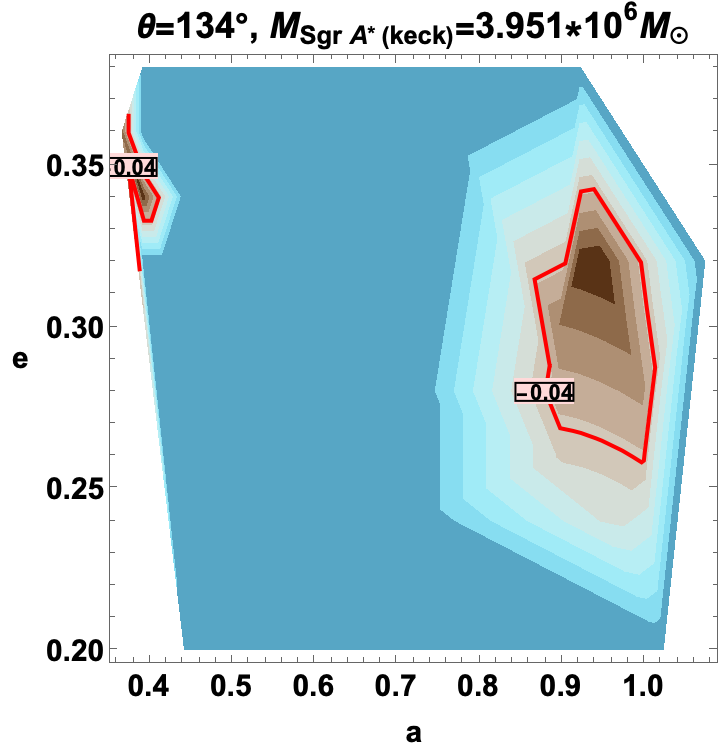}};
    \node[inner sep=0pt, text width=4cm ]  at (-6,8)
    {\includegraphics[width=1.5\textwidth]{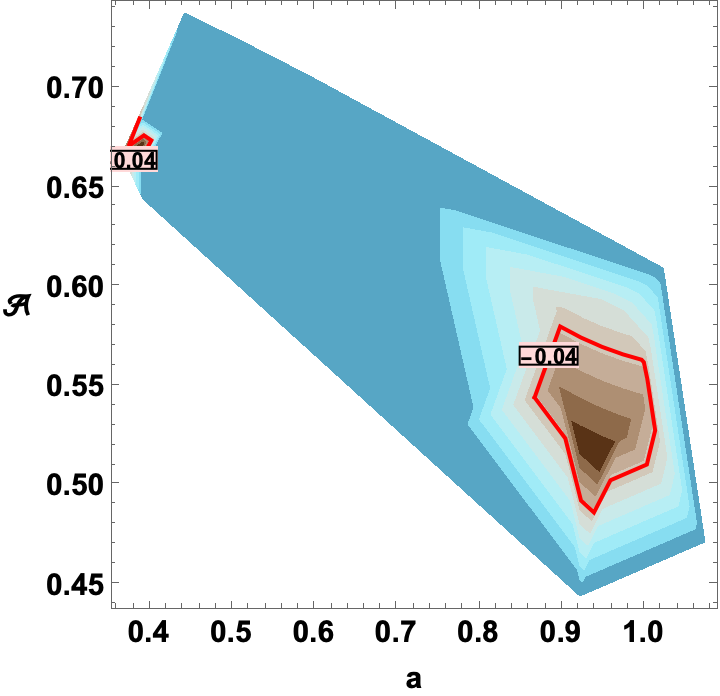}};
     \node[inner sep=0pt, text width=4cm ]  at (-6,2)
    {\includegraphics[width=1.5\textwidth]{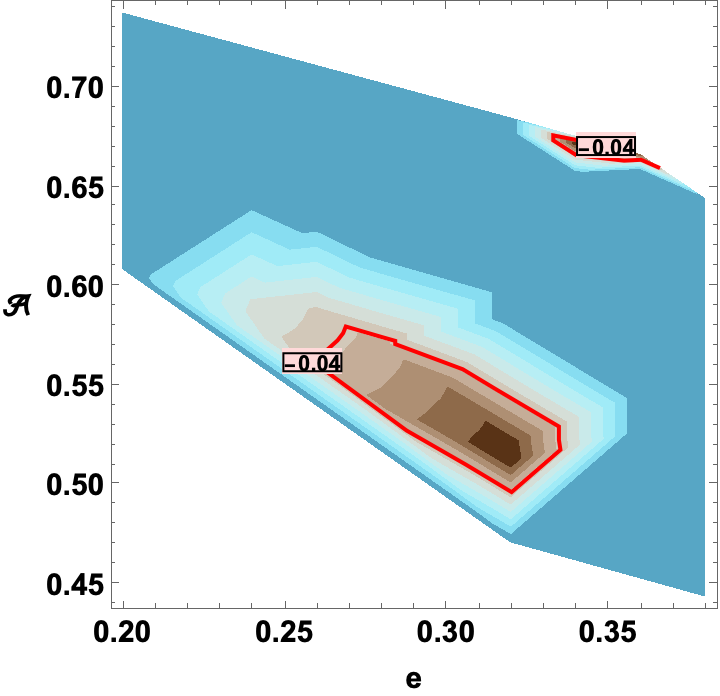}}; 

     \node[inner sep=0pt, text width=4cm ]  at (4,13.6) {\includegraphics[width=.4\textwidth]{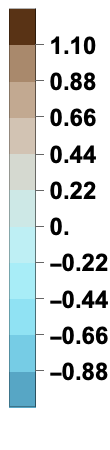}};
 \node[inner sep=0pt, text width=4cm ]  at (.5,13.3) {\includegraphics[width=.4\textwidth]{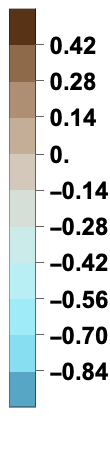}};
  \node[inner sep=0pt, text width=4cm ]  at (-1,-8.4) {\includegraphics[width=2\textwidth]{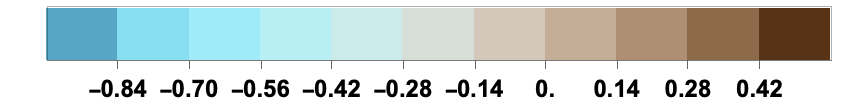}};

\end{tikzpicture}
\caption{\footnotesize  \it Contours illustrate the dependence of the fractional deviation $\bm \delta$ for $\textrm{M87}^\star$ and $\textrm{Sgr~A}^\star$ (VLTI \& Keck observatories) on different planes of the supersymmetric accelerating black hole parameters: charge $e$, rotation $a$, and acceleration $\mathcal{A}$. The solid red line represents the EHT measurements.
}
\label{fig3}
\end{figure*}

In Fig.\ref{fig2}, we present the contours of the angular diameter for the $\textrm{M87}^\star$ and $\textrm{Sgr~A}^\star$ black holes across different parameter planes: (charge, rotation), (acceleration, rotation), and (acceleration, charge). Meanwhile, Fig.\ref{fig3} illustrates the contours of the fractional deviation.

%\cleardoublepage

From both figures, one can notice that the observable $\bm{\delta}$ presents a behavior similar to that of the angular diameter $\mathcal{D}$. Furthermore,
 the supersymmetric black hole solution aligns well with the observational data of $\textrm{M87}^\star$ and $\textrm{Sgr~A}^\star$ illustrated by the red contours. This alignment suggests that the features of the black hole shadows, as observed by the EHT, exhibit characteristics consistent with supersymmetry predictions. The concordance between the theoretical models and empirical data highlights the potential presence of supersymmetry imprints in the shadows of these black holes, offering compelling evidence of supersymmetric phenomena in astrophysical observations. This stands in contrast to traditional particle physics experiments, which have so far failed to reveal such evidence.

%-------------------------------------------
The existence and nature of supersymmetry are pivotal open questions in physics, garnering significant attention, particularly in the context of quantizing gravity, unification theories, and string theory. Our conclusion, suggesting the first supersymmetric imprints from the EHT observations, presents an exciting avenue for further verification through additional observations.% such as light deflection experiments within our solar system and gravitational wave detections.

This new result underscores the burgeoning importance of black hole theory and phenomenology, especially given the rapid advancements in observational techniques. As these observations continue to develop, the study of black holes is poised to remain a vibrant and critical field of research, potentially offering deeper insights into the fundamental principles of our universe.%*****
%Our findings show 
%**************************************************************
%Imaging black holes has been established as an incredible tool since the first direct observation of the $\textrm{M87}^\star$ by the {\it EHT} collaboration. It has become so ubiquitous that humanity never bats an eye at its utility until physicists exclaim just how incredible it is that we manage to validate and constrain theories. What has become even more surprising is that the images of black holes have proven more and more useful even as our theories have become more complicated.
%**************************************************************
%Knowledge about the nature of the gravitational interaction in the near-horizon regime is expected to lead towards a better understanding of gravity and the nature of ultra-compact objects at a fundamental level
%
%
%GR has stood up to a variety of tests

%\section*{Data availability}
%Since this work is theoretical, there is no data used to support the findings of this study. 
\newpage
\bibliographystyle{apsrev4-1}%{rusnat}
\bibliography{Susy_shadow}
\end{document}